\documentclass[12pt, draftclsnofoot, onecolumn]{IEEEtran}
\usepackage{amsfonts}
\usepackage{amssymb}
\usepackage{amsthm}
\usepackage{amsmath,amsfonts,amssymb}
\usepackage[dvips]{graphicx}
\usepackage{verbatim}
\usepackage{bm}
\usepackage[ruled,vlined]{algorithm2e}
\usepackage{cite}
\usepackage{color}

\IEEEoverridecommandlockouts

\begin{document}
% paper title
\title{Millimeter-Wave Communication with Non-Orthogonal Multiple Access for 5G}

\author{Zhenyu Xiao,~
        Linglong Dai,~
        Zhiguo Ding,~
        Jinho Choi,~
        Pengfei Xia,~
and Xiang-Gen Xia
\thanks{Zhenyu Xiao is with the School of Electronic and Information Engineering, Beihang University, Beijing, P. R. China.}
\thanks{Linglong Dai is with the Department of Electronic Engineering, Tsinghua University, Beijing, P. R. China.}
\thanks{Zhiguo Ding is with the School of Computing and Communications, Lancaster University, LA1 4WA, UK.}
\thanks{Jinho Choi is with the School of Electrical Engineering and Computer Science, Gwangju Institute of Science and Technology (GIST), Korea.}
\thanks{Pengfei Xia is with the School of Electronics and Information Engineering, Tongji University, Shanghai, P. R. China.}
\thanks{Xiang-Gen Xia is with the Department of Electrical and Computer Engineering, University of Delaware, Newark, DE, USA.}
}
\vspace{-10mm}
% make the title area
\maketitle
\vspace{-10mm}
\begin{abstract}
To further improve the system capacity for 5G, we explore the integration of non-orthogonal multiple access (NOMA) in mmWave communications (mmWave-NOMA) for future 5G systems. Compared with the conventional NOMA, the distinguishing feature of mmWave-NOMA is that, it is usually characterized by transmit/receive beamforming with large antenna arrays. In this paper, we focus on the design challenges of mmWave-NOMA due to beamforming. Firstly, we study how beamforming affects the sum-rate performance of mmWave-NOMA, and find that with conventional single-beam forming, the performance may be offset by the relative angle between NOMA users. Then, we consider multi-beam forming for mmWave-NOMA, which is shown to be able to achieve promising performance enhancement as well as robustness. We further investigate the challenging joint design of the intertwined power allocation and user pairing for mmWave-NOMA. We also discuss the challenges and propose some potential solutions in detail. Finally, we consider hybrid spatial division multiple access (SDMA) and NOMA in mmWave communications, where some possible system configurations and the corresponding solutions are discussed to address the multi-user issues including multi-user precoding and multi-user interference (MUI) mitigation.
\end{abstract}


%\IEEEpeerreviewmaketitle
\section{Introduction}
%% mmWave for 5G is prominent
\IEEEPARstart{A}{s} the fifth generation (5G) wireless mobile communication becomes looming on the horizon, the requirements of 5G gradually become clearer, and among them the large aggregate capacity is one of the most critical issues~\cite{andrews2014will}. To realize high aggregate capacity, besides massive multiple-input multiple-output (MIMO) and ultra-dense network, millimeter-wave (mmWave) communication has been considered as one of the major candidate technologies \cite{andrews2014will,niu2015survey,rapp2013mmIEEEAccess}. Indeed, thanks to the abundant frequency spectrum resource, mmWave communication can provide a much higher capacity than the legacy low-frequency mobile communications working in the micro-wave band.

% rely on efficient MA , OMA is not enough for 5G, mmWave NOMA
When applying mmWave communication to the cellular systems, its benefit will highly rely on multiple access strategies. Subject to the limited radio resources, the existing time/frequen-cy/code division multiple access (TDMA/FDMA/CDMA) may face stringent challenges in supporting a greatly increased number of users in future 5G systems, which is supposed to connect massive users/devices \cite{andrews2014will}. Moreover, due to the obvious user diversity in 5G cellular systems, the data rate requirements will be quite different for different users. To allocate an equal radio resource to a user requiring a low date rate will be a waste. Such inefficient orthogonal multiple access (OMA), i.e., TDMA/FDMA/CDMA, may offset the benefit of improved capacity of mmWave communication in future 5G cellular systems. Different from these OMA schemes, the non-orthogonal multiple access (NOMA) strategy is able to support multiple users in the same (time/frequency/code) resource block realized by superposition coding in the power domain \cite{ding2014performance,saito2013non,Choi2014NOMA,Dai2015NOMA5G,Saito2013syslevl}. By exploiting corresponding successive interference cancellation (SIC) in the power domain at receivers, multiple users can be distinguished from each other, thus both the number of users and the spectrum efficiency can be increased.

The use of NOMA is also necessary for mmWave communication, since a larger bandwidth also calls for an efficient use of the acquired spectrum to support massive connectivity and the exponential traffic growth. On the other hand, the highly direction feature of mmWave propagation makes the users' channels (along the same or similar direction) highly correlated, which facilitates the integration of NOMA in mmWave communication, i.e., mmWave-NOMA \cite{Ding2017random,Daill2017}. An intrinsic difference between mmWave-NOMA and conventional NOMA is that, beamforming with a large antenna array is usually adopted in mmWave-NOMA. In \cite{Ding2017random}, random steering single-beam forming was adopted, which can work only in a special case that the NOMA users are close to each other. In \cite{Daill2017}, multi-beam forming was used to serve multiple NOMA users with arbitrary locations, but subject to lens antenna array (a low-complexity realization of hybrid precoding).

In this paper we focus on the design challenges of mmWave-NOMA due to beamforming using regular phased array. Specifically, we first study how beamforming affects the sum-rate performance of NOMA in Section II, and we find that with conventional single-beam forming, the sum-rate performance may be offset by the relative angle between the NOMA users. Then, we consider multi-beam forming for mmWave-NOMA, and show that mmWave-NOMA with multi-beam forming is able to achieve promising performance enhancement as well as robustness. However, the design of multi-beam forming is more challenging than conventional single-beam forming. In addition, different from conventional NOMA, beamforming usually intertwines with power allocation in mmWave-NOMA, and this issue is discussed in detail in Section III, where several potential solutions are proposed to solve the challenging joint design problem of power allocation and beamforming. Next, we consider the more challenging user pairing in mmWave-NOMA in Section IV, since it intertwines with both power allocation and beamforming. Finally, considering that it is not realistic in general to only use NOMA in mmWave communication, we investigate hybrid spatial division multiple access (SDMA) and NOMA in Section V, where the possible system configurations as well as corresponding solutions are discussed to address the multi-user issues, e.g., multi-user precoding and multi-user interference (MUI) mitigation.

%An immediate question comes that does mmWave-NOMA benefit over mmWave-OMA (e.g., mmWave-TDMA) or not? Although it has been recognized that NOMA generally benefits over TDMA in the legacy low-frequency cellular, in mmWave communications the situation is different. The localizations of users significantly affect the beam gain of the BS, and in turn affect the achievable rate of mmWave-NOMA. Hence, we study this issue in Section II, and we show that with conventional beamforming, the performance of mmWave-NOMA may be rather poor depending on the locations of the users. To sufficiently exploit the benefit of mmWave-NOMA, we further investigate smart beamforming for mmWave-NOMA in Section III, which can guarantee the performance of mmWave-NOMA. Next, we consider user management for mmWave-NOMA in Section IV, including a universal downlink/uplink framework, user pairing methods, and user joining issues. Last, we conclude the paper in Section V.

\section{Beamforming in mmWave-NOMA}
% mmWave beamforming,

\subsection{MmWave-NOMA with Conventional Single-Beam Forming}
In a conventional micro-wave cellular system with one antenna at the base station (BS), the transmission is isotropic, and beamforming is not involved for OMA or NOMA \cite{ding2014performance,saito2013non}. In contrast, mmWave transmission is directional instead of isotropic. In fact, beamforming with a large antenna array is a key characteristic of mmWave communication, which is used to compensate for the high path loss due to the high frequency. To serve multiple users, TDMA may be a preferred OMA for directional mmWave transmission, where the BS needs to perform beamforming to steer towards a single user in each time slot. Unlike mmWave-TDMA, mmWave-NOMA is able to serve multiple users in each time slot to support greater connectivity and increase the network capacity accordingly. Nevertheless, beamforming behaves differently in those two multiple access schemes, which is detailed as follows.

Firstly, it is noteworthy that in mmWave communication, subject to the hardware complexity and power consumption, beamforming is usually realized with some hardware constraints. For instance, the antenna weights are with the same constant-modulus (CM), and the number of radio-frequency (RF) chains is much smaller than that of the antennas \cite{xiao2016codebook}. Thus, mmWave beamforming is less flexible than MIMO beamforming. Subject to the hardware constraints, usually \emph{only a single beam} is formed (single-beam forming) in one time slot for existing analog beamforming designs, where wide beams are used for hierarchical search of the user direction in the initial beam alignment, and then narrow beams are used for data transmission \cite{rapp2013mmIEEEAccess,sun2014mimo,xiao2016codebook}. When the mmWave-TDMA BS serves only a single user in one time slot, beamforming is straightforward and a narrow beam can be easily formed to steer towards the user \cite{xiao2016codebook}. In contrast, when the mmWave-NOMA BS serves multiple users in one time slot, a narrow beam may probably not cover all users. Instead, a wide beam may be required to cover all served users in that time slot, and the beam width depends on the relative angle between these users. This may significantly reduce the beam gain and in turn offset the benefit of NOMA, because the beam gain is roughly inversely proportional to the beam width.

\begin{figure}[t]
\begin{center}
  \includegraphics[width=12 cm]{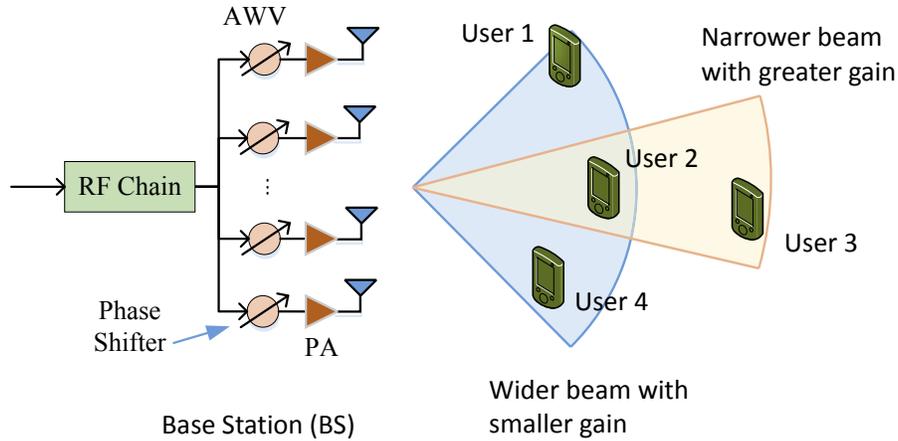}
  \caption{MmWave-NOMA with analog beamforming at the BS. The angle gap between two NOMA users affects the beam gain when single-beam forming is exploited. For instance, the beam gain is higher and the beam width is smaller when serving Users 2 and 3, because their angle gap is smaller than that of User 1 and User 4.}
  \label{fig:ConvBF}
\end{center}
\end{figure}

% wide beam, beam gain
To illustrate this issue, we give an example in Fig. \ref{fig:ConvBF}, where the mmWave-NOMA BS with analog beamforming (only one RF chain is used) needs to serve two users in one time slot. For analog beamforming, a large number of antennas share a single RF chain, and in the branch of each antenna, a phase shifter is used to only adjust the signal phase for analog beamforming \cite{sun2014mimo,xiao2016codebook}. It is obvious that the analog beamforming has a low hardware complexity, but it suffers from a low flexibility in beamforming, because only the phases of the antenna weights can be controlled. In this case, single-beam forming is usually considered \cite{rapp2013mmIEEEAccess,sun2014mimo,xiao2016codebook}. When serving Users 2 and 3, the BS only needs to form a narrow beam, because the angle gap between Users 2 and 3 is small. In such a case, the beam gain will be high. However, when serving Users 1 and 4, the BS has to form a much wider beam, because the angle gap between Users 1 and 4 is much larger. As a result, the beam gain of the BS will be much lower, which degrades the performance of NOMA.

\begin{figure}[t]
\begin{center}
  \includegraphics[width=12 cm]{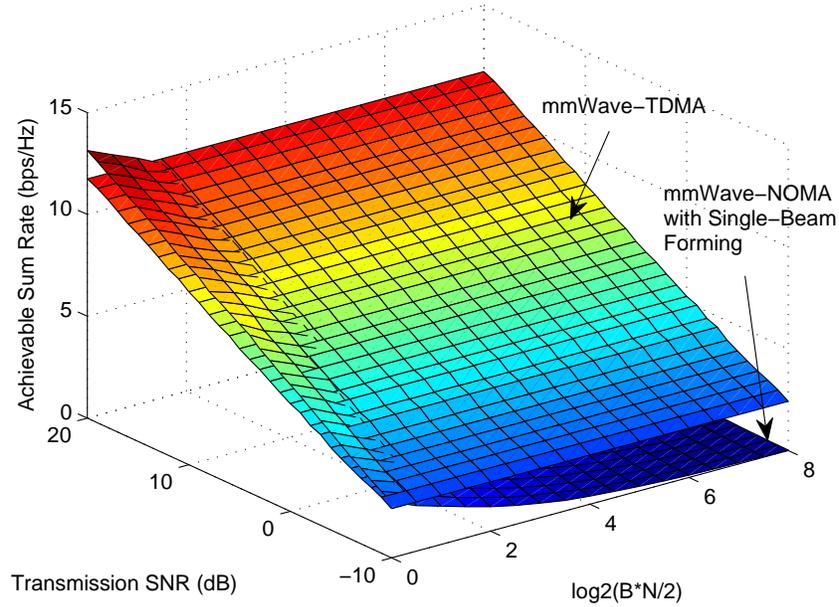}
  \caption{Performance comparison between mmWave-NOMA with single-beam forming and mmWave-TDMA, where $N=32$ is the number of antennas at the BS. Transmission SNR is the SNR without considering the beam gain at the BS. $B$ is the required beam width to cover the two NOMA users. The mmWave channel model in \cite{xiao2016codebook} and \cite{xiao2017codebook} is adopted between the BS and a user, and the number of paths is set to 1 for simplicity. This setting is also adopted in the other simulations in this paper.}
  \label{fig:SAR_B}
\end{center}
\end{figure}

% beamforming in mmWave-NOMA performance
We compare the performances of mmWave-NOMA with single-beam forming and mmWave-TDMA by assuming a typical two-user case with $N=32$ antennas at the BS. According to \cite{xiao2016codebook}, the narrowest beam width for the array with $N$ elements is roughly $2/N$ in the cosine angle domain when steering to a user, and the beam gain can be roughly computed as $2/B$, where $B$ is the beam width. For mmWave-TDMA, the beam gain is roughly $N$, as a narrow beam can be shaped to steer towards a user in each time slot. For mmWave-NOMA, the beam gain varies as $B$. We assume the average channel power of User 1 is 6 dB higher than that of User 2, and in mmWave-NOMA the stronger user and weaker user are allocated with $1/4$ and $3/4$ of the total transmission power, respectively. Fig. \ref{fig:SAR_B} shows the performance comparison in terms of sum achievable rate. We can find that when the required beam width to cover the two users is $B=2/N$, mmWave-NOMA outperforms mmWave-TDMA due to the higher achievable rate of NOMA than TDMA. However, as the required beam width becomes larger, the performance of mmWave-NOMA deteriorates, and even becomes worse than that of mmWave-TDMA for a large $B$, since the beam gain is significantly reduced in this case.

% summary
In summary, the directional mmWave transmission makes mmWave-NOMA quite different from micro-wave NOMA, i.e., the performance of mmWave-NOMA with single-beam forming may be significantly affected by the relative locations of NOMA users. In particular, the beam gain of the BS will be lower when the angle of the two users is larger, which may offset the benefit of mmWave-NOMA.

\subsection{MmWave-NOMA with Multi-Beam Forming}
As mmWave-NOMA with single-beam forming does not behave efficiently and robustly enough \cite{Ding2017random}, we consider mmWave-NOMA with multi-beam forming, which means that the BS can form \emph{multiple narrow beams} to steer towards multiple NOMA users simultaneously \cite{Daill2017}. It should be emphasized again that although it may be a natural choice to shape multiple narrow beams for multiple users, multi-beam forming under the hardware constraint of analog beamforming is seldom considered in mmWave communications to the best of our knowledge, since it is challenging to form \emph{multiple} narrow beams by using analog beamforming with only \emph{one} RF chain. Besides, the concept of multi-beam forming here is different from that in \cite{Daill2017}, where $M$ RF chains rather than a single RF chain are used to form $M$ narrow beams.

\begin{figure}[t]
\begin{center}
  \includegraphics[width=13 cm]{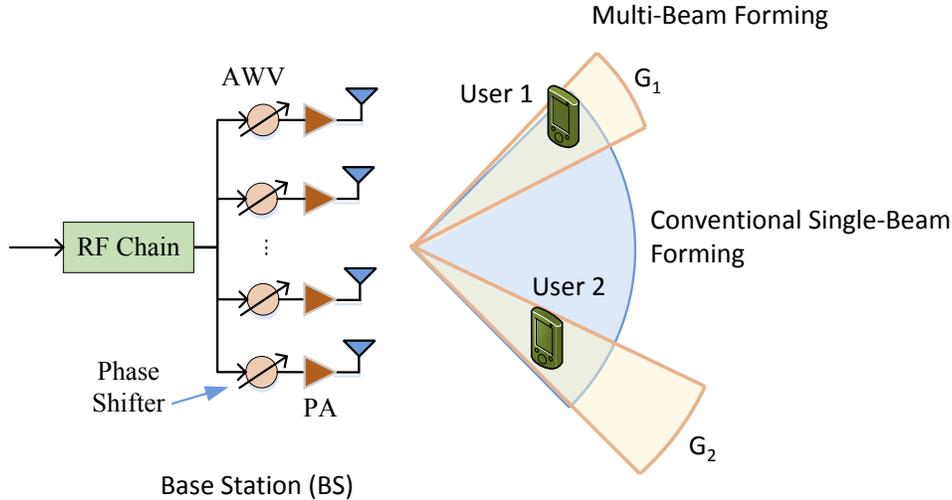}
  \caption{Comparison between single-beam forming and multi-beam forming for mmWave-NOMA.}
  \label{fig:SmartBF}
\end{center}
\end{figure}

As shown in Fig. \ref{fig:SmartBF}, the BS with multi-beam forming forms two narrow beams to cover two NOMA users (User 1 and User 2) in the same time slot. Intuitively, since multi-beam forming covers a narrower range than single-beam forming, a higher beam gain can be achieved, and such achieved beam gain will not be reduced even when the angles of two users become larger. More importantly, with multi-beam forming, the beam gains for different NOMA users can be different according to their channel qualities. For instance, in Fig. \ref{fig:SmartBF}, the beam gains $G_1$ for User 1 and $G_2$ for User 2 can be different. This feature is very important for mmWave-NOMA, because in addition to the degree of freedom in the power domain, it provides another degree of freedom, i.e., beamforming, to improve the performance of mmWave-NOMA, which will be discussed in detail later.

\begin{figure}
\begin{minipage}[t]{0.5\linewidth}
\centering
\includegraphics[width=9.0 cm]{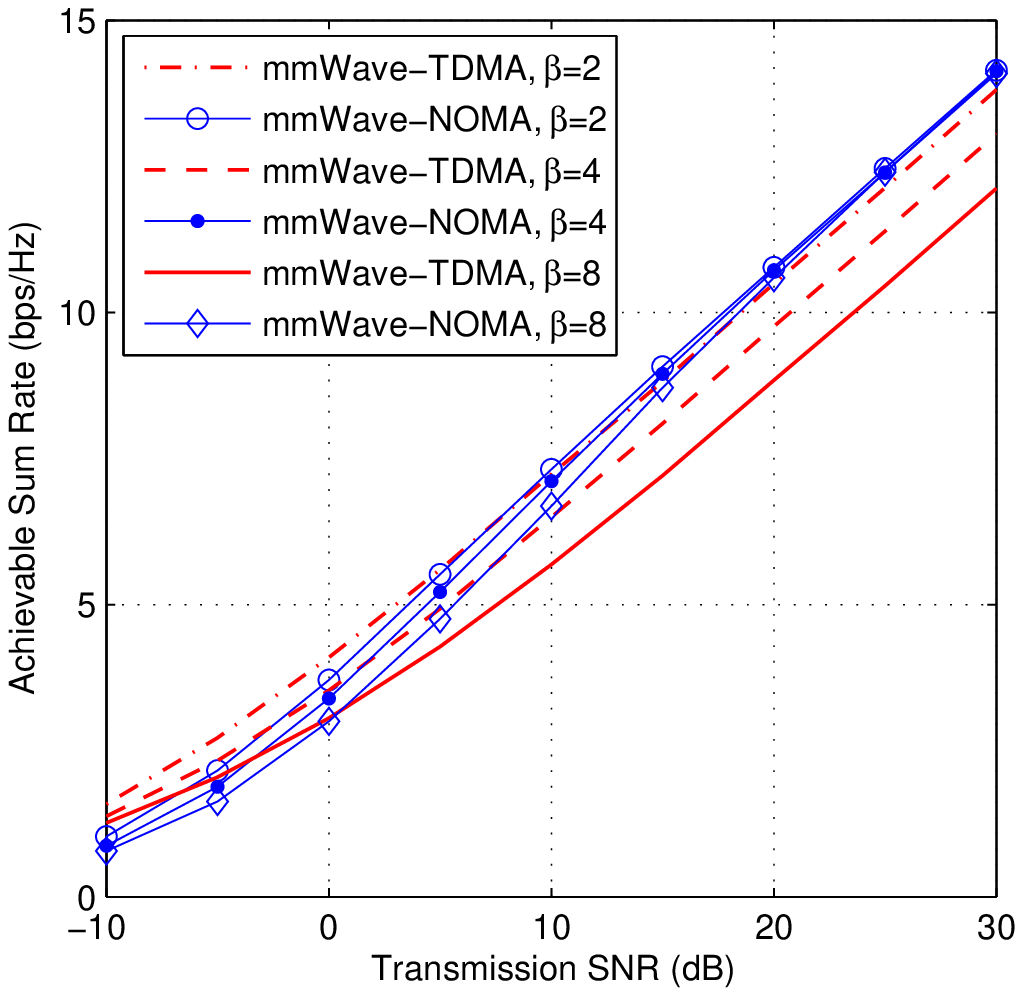}
\end{minipage}
% vfill
\begin{minipage}[t]{0.5\linewidth}
\centering
\includegraphics[width=9.0 cm]{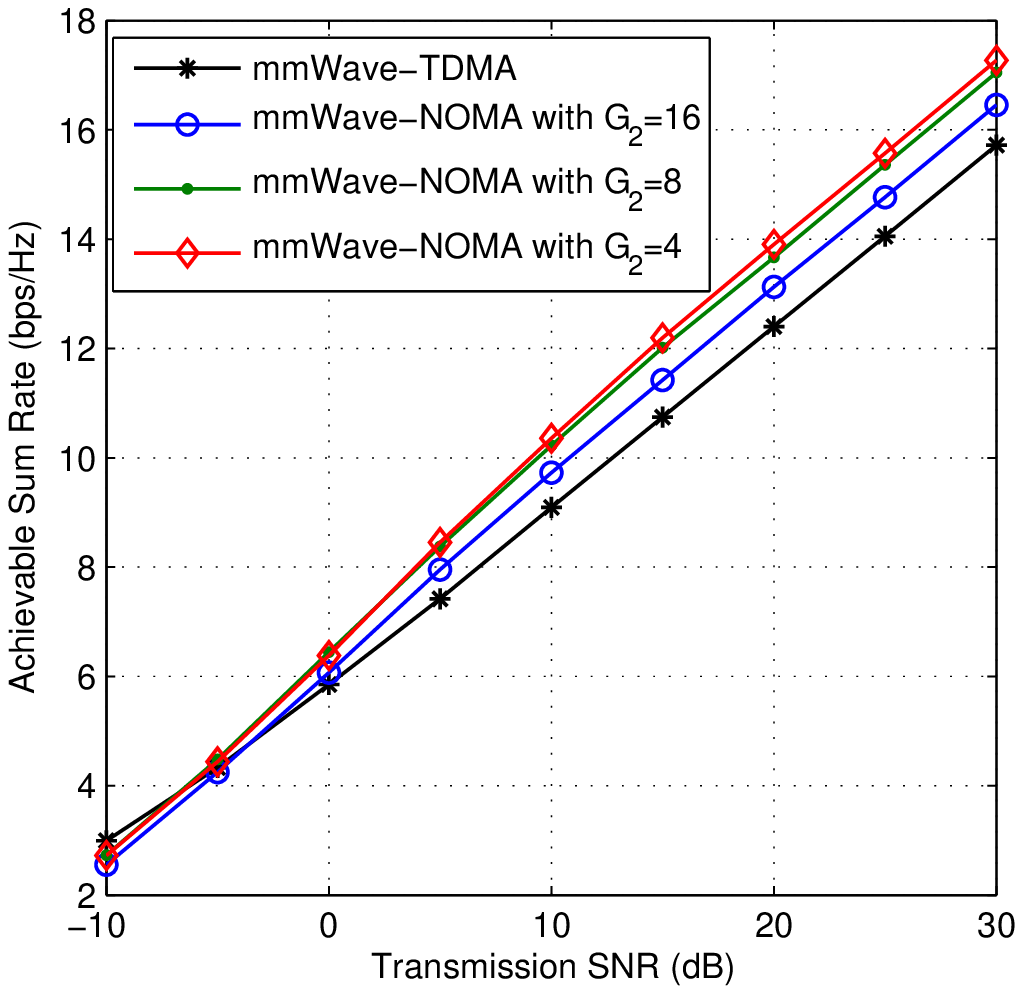}
\end{minipage}
\caption{(a): Performance comparison between mmWave-NOMA with multi-beam forming and mmWave-TDMA. $\beta$ is the ratio of the average channel gain of User 1 to that of User 2. For mmWave-NOMA with multi-beam forming, $G_1=G_2=16$. (b): Effect of setting different beam gains for different users in mmWave-NOMA with multi-beam forming. The ratio of the average channel gain of User 1 to that of User 2 is $\beta=3$. When $G_2$ is smaller, $G_1$ is greater, and the difference of channel gains between the two users is more significant.}
\label{fig:smartBF_per}
\end{figure}

We compare the sum-rate performance of mmWave-NOMA with multi-beam forming with that of mmWave-TDMA in Fig. \ref{fig:smartBF_per} (a), where $\beta$ is the ratio of the average channel gain of User 1 to that of User 2. In the comparison, $G_1=G_2=16$ for mmWave-NOMA with multi-beam forming, i.e., we do not enlarge the difference of channel gains by setting different beam gains for these two users. The total average channel power is identical for both mmWave-NOMA and mmWave-TDMA for fair comparison. We can observe that with multi-beam forming, mmWave-NOMA performs better than mmWave-TDMA in general. In addition, as $\beta$ becomes larger, the superiority of mmWave-NOMA becomes more significant.

As aforementioned, by appropriately setting the beam gains for different users, the performance of mmWave-NOMA can be further improved. We show this in Fig. \ref{fig:smartBF_per} (b), where we assume the beam widths of User 1 and User 2 are the same, i.e., $2/N$ in the cosine angle domain; hence we have $(G_1+G_2)\times 2/N=2$ or $G_1+G_2=N$. The ratio of the average channel gain of User 1 to that of User 2 is $\beta=3$. When $G_2$ is smaller, $G_1$ becomes larger, and the channel gain difference between these two users is more significant. We can find that as $G_2$ becomes smaller, the sum-rate performance of mmWave-NOMA is improved. It is noteworthy that the improvement becomes slower as $G_2$ becomes smaller. When $G_2$ is small enough, or the difference of the channel plus beam gains of these two users is large enough, further reducing $G_2$ and increasing $G_1$ do not help much to improve the sum-rate performance.

\subsection{Challenge of Multi-Beam Forming}
Compared with single-beam forming, multi-beam forming has distinctive advantages for mmWave-NOMA. However, the antenna weight vector (AWV) design is more challenging for multi-beam forming. For single-beam forming, only one beam needs to be formed. In comparison, for multi-beam forming, multiple beams with different beam gains need to be formed. Considering that the AWV is subject to the CM constraint due to phase shifters, the AWV design is difficult.

%In most of cases, many wide beams can be pre-designed and stored \cite{alkhateeb2014channel,xiao2016codebook}.
% sub-array technique
One possible way to design AWV for multi-beam forming is the sub-array technique \cite{xiao2016codebook,xiao2017codebook}. As we need to form multiple beams, a natural method is to divide a large antenna array into several sub-arrays, and let different sub-arrays steer towards different directions. For instance, if we divide a large array into 2 sub-arrays with CM AWVs ${\bf{w}}_1$ and ${\bf{w}}_2$, respectively, it is easy to design ${\bf{w}}_1$ to steer towards a direction, while ${\bf{w}}_2$ is designed to steer towards another direction \cite{xiao2016codebook,xiao2017codebook}. Thus, the overall AWV ${\bf{w}}=[{\bf{w}}_1^{\rm{T}},{\bf{w}}_2^{\rm{T}}]^{\rm{T}}$ can steer towards two different directions. Note that such direct combination may lead to significant sidelobes, and it is better to use an adjustable coefficient for each sub-array to minimize the sidelobes, i.e., to let ${\bf{w}}=[e^{j\theta_1}{\bf{w}}_1^{\rm{T}},e^{j\theta_2}{\bf{w}}_2^{\rm{T}}]^{\rm{T}}$, where $e^{j\theta_1}$ and $e^{j\theta_2}$ are coefficients to minimize the sidelobes \cite{xiao2016codebook,xiao2017codebook}. It is noteworthy that when the beam gains of the multiple beams are the same, the sub-array technique is suitable for multi-beam forming. However, when the beam gains are different, the numbers of antennas of the sub-arrays need to be carefully designed to fulfil the gain requirements under the CM constraint.

%\begin{figure}[t]
%\begin{center}
%  \includegraphics[width=16 cm]{Fig3.eps}
%  \caption{Two candidates to divide a large antenna array to sub-arrays.}
%  \label{fig:Fig3}
%\end{center}
%\end{figure}

% Optimization technique
Another possible way to design the CM AWV for multi-beam forming is to apply the optimization approach. Since the number of antennas is large in general, the dimension (the number of variables) of the established optimization problem will be large. As a result, the formulation of a tractable optimization problem is critical. One challenging issue in the problem formulation is how to deal with the CM constraint and the user gain constraints, which are all non-convex. For instance, if we want to design ${\bf{w}}$ to form two different beams with gains $G_1$ and $G_2$, respectively, it is natural to minimize $||{\bf{w}}||^2$ (or $\alpha$) subject to the CM constraint $|{\bf{w}}|=\alpha{\bf{1}}$ and gain constrains $|{\bf{w}}^{\rm{H}}{\bf{a}}(\phi_i)|=G_i$, where $i=1,2$ and ${\bf{a}}(\phi_i)$ is a given steering vector towards the direction $\phi_i$ \cite{xiao2016codebook}. However, with these equality constraints, the problem is generally difficult to solve, not only because they are non-convex constraints, but also because the equality constraints are usually too strict to find an appropriate CM AWV. In such a case, some relaxation may be induced to ease the problem. For instance, we may relax the CM constraints to minimize the maximal absolute weight of the antenna weights, i.e., to minimize $\alpha$ subject to $|{\bf{w}}|\prec \alpha{\bf{1}}$ where $\prec$ is componentwise inequality. Meanwhile, we may relax the beam gain requirements from equality to inequality, i.e., $|{\bf{w}}^{\rm{H}}{\bf{a}}(\phi_i)|\geq G_i$. With these relaxations the problem usually becomes easier to solve, but the CM constraint is not necessarily satisfied via solving the relaxed problem. In fact, the target behind the relaxed problem is to let each absolute weight of the AWV be the same while satisfying the gain constraints. Hence, after solving the relaxed problem, we need to normalize the obtained AWV to satisfy the CM constraint with the phases of its elements unchanged.

\section{Joint Power Allocation and Beamforming}
In conventional single-antenna NOMA systems, power allocation plays a critical role in satisfying the performance requirements, e.g., maximizing the achievable sum rates of all the NOMA users under user rate constraints, or maximizing the minimal rates of the NOMA users. Taking sum rate maximization as an example, a user power can be seen as a degree of freedom to tune user rates so as to maximize the sum rate. In mmWave-NOMA systems, as we has discussed above, in addition to user power, beamforming can be a new degree of freedom to tune the user performances. Specifically, the effective channel gains of the users can be changed by beamforming. For instance, considering the 2-user mmWave-NOMA system in Fig. \ref{fig:SmartBF}, where the channel gains and beam gains for these two users are $h_1$, $G_1$ and $h_2$, $G_2$, respectively, i.e., the effective channel gains of the two users are $h_1G_1$ and $h_2G_2$, respectively. Thus, by changing the user gains via beamforming, i.e., $G_1$ and $G_2$, the effective channel gains can be changed accordingly.

Now we have two degrees of freedom to improve the performance, i.e. power allocation and beamforming. Given fixed beamforming, power allocation in mmWave-NOMA will be the same as that in the conventional single-antenna NOMA system. While given fixed power allocation, beamforming can be realized by using the sub-array technique or the optimization approaches introduced in the previous section. However, in most cases, power allocation intertwines with beamforming in mmWave-NOMA, because the achievable rates of the users depend on both power allocation and beamforming. As a result, we usually need to consider the joint power allocation and beamforming problem, i.e., we need to find optimal power allocation for each user as well as optimal beamforming vector subject to the CM constraint at the BS.

For example, we can also consider the 2-user mmWave-NOMA system in Fig. \ref{fig:SmartBF}, where the users are equipped with a single antenna. A problem is how to maximize the achievable sum rate of the two users provided that the channel is known \emph{a priori}. It is clear that if there are no minimal rate constraints for these two users, the achievable sum rate can be maximized by allocating all the power to User 2 and meanwhile beamforming towards User 2, whose channel gain is better. However, when there are minimal rate constraints for these two users, the power allocation intertwines with the beamforming design, which makes the problem complicated. Similar challenges applied to downlink/uplink transmission with the target of maximizing the sum rate or maximizing the minimal user rate.

As this kind of problem is non-convex and may not be converted to a convex problem with simple manipulations, it may be infeasible to make use of the existing optimization tools. On the other hand, to directly search the optimal solution is computationally prohibitive because the number of variables is large in general. A potential solution is to decompose the original joint power allocation and beamforming problem into two sub-problems: one is a power and beam gain allocation problem, and the other is a beamforming problem under the CM constraint, i.e., to determine $\{P_1,~P_2,~G_1,~G_2\}$ first, where $\{P_1,~P_2\}$ are powers for these two users, and then determine their beamforming vectors using the approaches introduced in the previous section. Although the original problem is difficult to solve, the two sub-problems are relatively easy to solve, and thus we can obtain a sub-optimal solution. In addition to this solution, alternating optimization can also be used to find a solution, i.e., alternatively optimize the power allocation with a fixed beamforming vector and the beamforming vector with fixed user powers.

In addition to the optimization method, some intuitive approaches with lower complexity can also be adopted to find a solution for the joint power allocation and beamforming problem. For instance, to maximize the sum rate, most power or beam gain should be allocated to User 2, which has the better channel quality, while only necessary power or beam gain should be allocated to User 1 to satisfy the rate constraint. For User 1, although its achievable rate increases with $P_1$ and $G_1$, increasing $P_1$ is more efficient to improve its rate, because increasing $G_1$ also increases the multi-user interference, i.e., the signal of User 2, while increasing $P_1$ reduces the multi-user interference on the contrary. Using these intuitive observations, we may set appropriate powers and beam gains for the two users, and then further determine the beamforming vector.

In brief, joint power allocation and beamforming is a key problem in mmWave-NOMA, which calls for extensive further studies.

\section{User Pairing in mmWave-NOMA}
% review单天线user paring
In conventional NOMA, user pairing is used to enable hybrid multiple access \cite{Ding2016UserPairing,Mei2016JointUP_PA}, in which NOMA is combined with the OMA schemes. Specifically, user pairing decides how to divide the users into multiple groups, where NOMA is implemented within each group and different groups are allocated with orthogonal radio resources. Obviously, the performance of the hybrid multiple access scheme is highly dependent on user pairing. Generally, it is difficult to find the optimal user pairing strategy \cite{Ding2016UserPairing,Mei2016JointUP_PA} due to its enumeration feature, except applying exhaustive search. The problem is that, when the number of users is large, the computational complexity of exhaustive search would be prohibitively high. Thus, intuitive approaches with lower complexity can be adopted, i.e., weaker users are usually paired with strong users, as the benefit of NOMA will increase as the channel gain difference \cite{Ding2016UserPairing,Mei2016JointUP_PA}. Nevertheless, even with the intuitive approaches, the challenge of user pairing is usually intensified by user power allocation, which is jointly designed with user pairing in general \cite{Ding2016UserPairing,Mei2016JointUP_PA}.

\begin{figure}[t]
\begin{center}
  \includegraphics[width=15 cm]{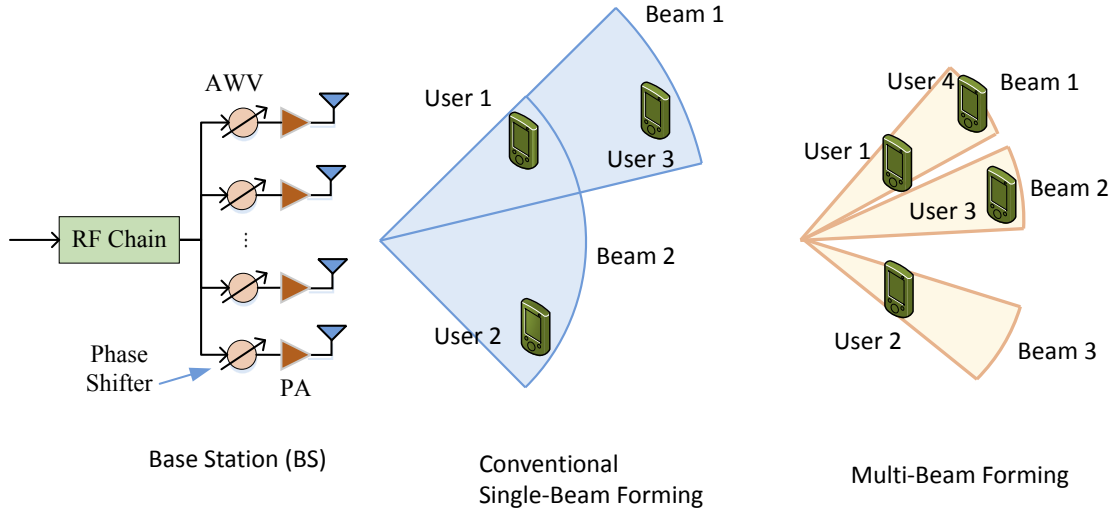}
  \caption{Beamforming affects user pairing in mmWave-NOMA.}
  \label{fig:user_pairing}
\end{center}
\end{figure}

% conventional beamforming 的user paring
In mmWave-NOMA, user pairing also faces similar challenges as those in the conventional NOMA systems, including the enumeration feature as well as entangling with power allocation. A new issue that also affects user pairing in mmWave-NOMA is beamforming. In particular, the channel gains of the users are the main factors to affect user pairing in conventional NOMA. However, in mmWave-NOMA the relative angles between the users also affect user pairing, because they affect the beam gains. We take the mmWave-NOMA system shown in Fig. \ref{fig:user_pairing} for instance, where the middle sub-figure shows the situation with conventional single-beam forming, while the right sub-figure shows the situation with multi-beam forming. We first see the middle sub-figure, where we need to select two users as a NOMA group. Clearly, User 3 can be paired with either User 1 or User 2 if only channel gain is considered. However, when using conventional single-beam forming in mmWave-NOMA, a wide beam, i.e., Beam 2, needs to be formed to cover the group of Users 3 and 2, because the relative angle between them is large. In contrast, a narrower beam, i.e., Beam 1, needs to be formed to cover the group of Users 3 and 1, because the relative angle between them is smaller. As a result, the achievable beam gain when pairing Users 3 and 1 is higher than pairing Users 3 and 2.

The situation is different when using multi-beam forming in mmWave-NOMA, as shown in the right sub-figure of Fig. \ref{fig:user_pairing}. With multi-beam forming, two narrow beams, rather than one single wide beam with conventional single-beam forming, are formed to cover the two NOMA users. In such a case, it is almost the same to pair Users 3 and 2 as to pair Users 3 and 1, provided that the channel gains of Users 1 and 2 are similar, because it is the same with multi-beam forming to cover Users 3 and 2 with Beams 2 and 3 as to cover Users 3 and 1 with Beams 2 and 1. However, when the relative angle between User 3 and 1 is very small, e.g., smaller than $2/N$ in the cosine angle domain, the BS may not form two different beams to cover them, because the smallest beam width is $2/N$  \cite{xiao2016codebook}. Instead, the BS may form only one narrow beam to cover both users. In such a case, both users can achieve a higher beam gain. In particular, with multi-beam forming to cover Users 4 and 1 can achieve a higher beam gain than to cover Users 4 and 2, because the former only needs one single narrow beam, while the latter needs two narrow beams.

In summary, user pairing in mmWave-NOMA is more challenging than that in conventional NOMA, which intertwines both power allocation and beamforming. Substantial research is needed to reveal the relationship between user pairing, power allocation and beamforming, as well as the interplay between them.

% multi-beam forming的user paring

%配一幅图

\section{Hybrid SDMA and NOMA}

% 2个RF服务4个用户，与NOMA及BDMA区别
%[Isn't SDMA a type of NOMA already? Or do you mean that NOMA only applies to non-orthogonal in Time/Frequency/Code? We need to clarify the difference. Otherwise, people will get confused.]
%
%[add a reference to my TCOM paper in 2016 with R. Heath.]

\begin{figure}[t]
\begin{center}
  \includegraphics[width=15 cm]{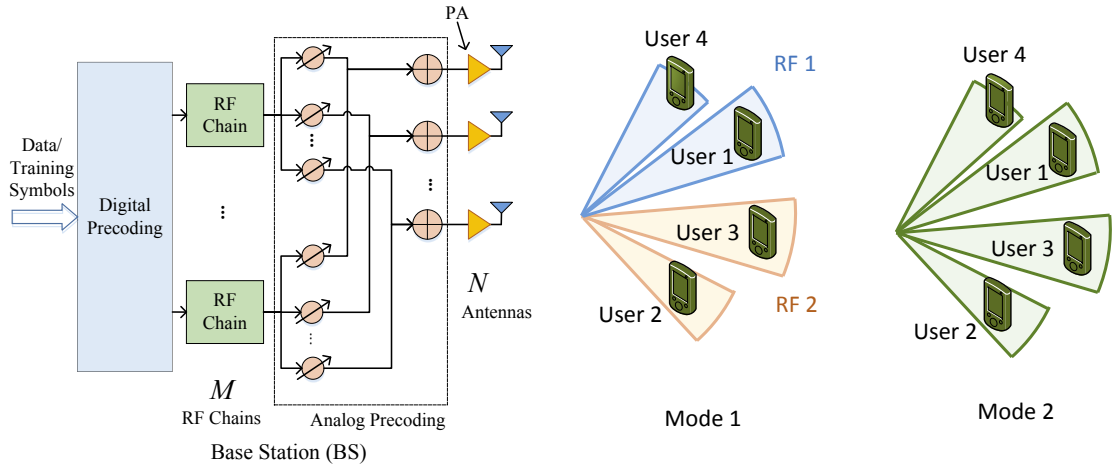}
  \caption{System structure of hybrid SDMA and NOMA in mmWave communications.}
  \label{fig:hybrid}
\end{center}
\end{figure}

In conventional mmWave communications, a BS can only serve one user when applying analog beamforming in one time/frequency resource block, because there is only one RF chain. To serve multiple users with SDMA, a BS needs to exploit hybrid analog/digital beamforming \cite{Daill2017}, as shown in Fig. \ref{fig:hybrid}. The hybrid structure has $M$ RF chains, and thus can support at most $M$ users by using SDMA. Thanks to the NOMA technology, multiple users now can be served in the same resource block with analog beamforming as introduced in Section II. However, the number of NOMA users is limited in general, because the achievable rates of the users with weak channel gains decrease as the number of NOMA users due to MUI. In such a case, a potential method to increase the number of users in mmWave-NOMA is to use hybrid SDMA and NOMA. In particular, the BS may have a hybrid beamforming structure with $M$ RF chains, and exploit NOMA with each RF chain and SDMA between different RF chains. If each RF chain can serve $K$ NOMA users, the hybrid SDMA and NOMA strategy can support at most $MK$ users.

%模式一，一个RF：2/2，但是有MUI，beamforming design
We show an example for the hybrid SDMA and NOMA strategy. Suppose $M=K=2$ in Fig. \ref{fig:hybrid}, i.e., there are two RF chains and each RF chain supports two NOMA users. We call this configuration Mode 1, as shown in the middle sub-figure. The BS sets beamforming vector of the first RF chain, i.e., RF 1, to form two narrow beams to cover Users 4 and 1, belonging to a NOMA group. In this process, multi-beam forming is adopted. Meanwhile, the BS sets beamforming vectors of the second RF chain, i.e., RF 2, to form another two narrow beams to cover Users 2 and 3, belonging to another NOMA group. Power allocation is performed among only NOMA users corresponding to the same RF chain, e.g., among Users 4 and 1 or Users 2 and 3. When designing the beamforming vector for each RF chain and power allocation for the NOMA users, the methods introduced in Sections II and III may be used. Moreover, here we also face the user pairing problem, and the relevant considerations in Section IV are applicable.

It is noteworthy that the above manipulations have implicitly ignored the MUI from other NOMA groups. This is reasonable when the number of RF chains is small in mmWave communications, because beams are designed to precisely steer to the users, such that the MUI from other NOMA groups is negligible. However, when the number of RF chains is not so small, the MUI from other NOMA groups needs to be considered in the designs of beamforming, power allocation and user pairing. In such a case, the designs for one RF chain are entangled with those for other RF chains, and the above manipulations may not be applicable.

%模式二，统一，当成4个，相当于过载SDMA，beamforming design
For the case that the MUI cannot be ignored, we propose the configuration of Mode 2, as shown in the right sub-figure of Fig. \ref{fig:hybrid}. In this mode, all RF chains jointly serve all users. In particular, the BS forms four narrow beams to cover the four users using the hybrid beamforming structure, and power allocation is performed among the four users. In fact, Mode 2 can be seen as either an overloaded SDMA mmWave communication system, where the number of users is larger than that of the RF chains, or an mmWave-NOMA system, where a hybrid structure is adopted for beamforming. Compared with the beamforming and power allocation in analog-beamforming mmWave-NOMA, the digital-domain processing adds a new degree of freedom to optimize the system performance. To be specific, in analog-beamforming mmWave-NOMA, we design analog beamforming and power allocation, while in hybrid-beamforming mmWave-NOMA, we need to design digital precoding, analog precoding and power allocation.

In brief, when MUI cannot be ignored, the relevant designs for hybrid-beamforming mmWave-NOMA can be rather challenging. Further studies are needed to find promising solutions for hybrid-beamforming mmWave-NOMA.

\section{Conclusions}
In this paper, we investigated many design challenges on the beamforming issues of mmWave-NOMA for the future 5G system. We first showed that with conventional single-beam forming, the sum-rate performance of NOMA may be offset by the angular separation between the NOMA users. We then discussed multi-beam forming for mmWave-NOMA, which is shown to be able to achieve improved sum-rate performance and robustness. Meanwhile, we showed that the design of multi-beam forming is more challenging than single-beam forming, and the sub-array and optimization techniques would be applicable. As for mmWave-NOMA beamforming usually intertwines with power allocation, the formulation of an optimization problem was shown to be critical for joint power allocation and beamforming design. Problem decomposition, alternating optimization, and some intuitive methods may help to find sub-optimal solutions. A more challenging issue in mmWave-NOMA is user pairing. Besides power allocation, we showed that beamforming also affects user pairing. Substantial research is in demand to reveal the relationship between user pairing, power allocation and beamforming. Finally, for hybrid SDMA and NOMA, it was shown that the strength of MUI determines the system configuration. When MUI can be ignored, the system would be configured as multiple independent analog-beamforming mmWave-NOMA, such that the strategies for analog-beamforming mmWave-NOMA are applicable. While when MUI cannot be ignored, the system would be configured as an overloaded SDMA structure or mmWave-NOMA with a hybrid beamforming structure, where we need to jointly design digital precoding, analog precoding and power allocation to mitigate MUI and optimize the performance.

%\bibliographystyle{IEEEtran} % use IEEEtran.bst style
%\bibliography{IEEEabrv,Xiao60GHz,Xiao5GnNOMA}

\end{document}